\documentclass[12pt, a4]{article}
\usepackage{graphicx}
\begin{document}

\title{Thermo-acoustics and its detection in a premixed flame}
\maketitle

\author{Ratan Joarder$^a$,  Siba P. Choudhury$^a$, Syam S.$^a$, Nagendra  Singh$^b$, S. K. Biswas$^b$}
\newline
\newline
$^a$Aerospace Engineering Department, Indian Institute of Technology Kharagpur, West Bengal-721302, India \\$^b$Physical Sciences Department, Indian Institute of Science Education and Research Mohali, Knowledge city, Sector 81, SAS Nagar, Manauli PO 140306, India

\textit{jratan@aero.iitkgp.ac.in}

\begin{abstract}
A new optical technique based on light-matter interaction is devised in-house to detect thermo-acoustic disturbances generated after ignition and during propagation of a premixed flame front in a half open channel. The technique involves passing a polarized laser light through a medium whose density or refractive index varies due to the passage of acoustic waves and/or flame front and then capturing the leaked depolarised light through an analyser by a photo-detector. The technique is applied to combustor involving premixed flame propagation and tulip inversion. The thermo-acoustic signals and the flame front are distinguished by comparing the oscilloscope signal with high speed photography of the flow-field. Acoustic waves are found to intercept the flame propagation at various axial locations and time instants.
\end{abstract}


\section{Introduction}
Thermo-acoustic disturbances in gas turbine combustors or any other combustors per se could be lethal if not controlled properly. The heat release rate and combustor’s inherent acoustics could work together in a feed-back loop to raise the pressure inside the combustor by order of magnitude and cause material damage to the combustor. Of late, an important practice in combustor design is to characterise its thermo-acoustic properties before it is placed into service. Hence, detection and measurement of the thermo-acoustic signal are important steps of the combustor design process. Conventionally, wall mounted transduces are used to measure the pressure. However, when the need is to detect the signal in the bulk flow, the technique becomes inefficient due to the physical dimension of the transducers, as it can affect the flow-field properties. The high temperature (~1600 K) of the bulk flow is also a matter of concern in their use. An optical technique could solve the above two if devised properly. Also, the invisible acoustics can influence the flame characteristics which can alter the heat release rate and hence the combustion efficiency.
   A classic example where thermo-acoustic disturbances and a moving flame front are present simultaneously and possibly interact among themselves is tulip flame \cite{ELL, SAL, DUN} in half open or closed channels within specific aspect ratio range. The reasons behind tulip inversion of the flame has been investigated over decades. Although, there is a strong opinion that the sudden reduction in flame surface area after the flame skirt touches the walls of the chamber causes the retardation of the flame velocity and subsequent inversion of the flame front, the reason behind the typical shape of the flame front after the inversion is not understood yet. Flame instability theories (Darrieus-Landau, hydrodynamic etc.) are also invoked to further consolidate the above opinion. Numerical simulations \cite{HUA, XIA} however, indicate the presence of strong acoustic waves right after the ignition and its subsequent interaction with the flame-front and the flow-field. To the authors’ knowledge, an experimental detection of the acoustic waves in the context of tulip inversion is not reported in the open literature. In the present work an optical technique, based on the principles of polarization and depolarization effect under lighter-matter interaction phenomenon, is devised in-house in order to capture the acoustic waves along with the moving flame front in an existing experimental facility \cite{SIBA}. The principles of acoustic wave and flame front detection are described in the following section. 
\section{The principle of acoustic wave and flame front detection}
Light is known as visible spectra of electromagnetic wave that travels at certain speed in vacuum.  However, when light travels through a medium other than vacuum, it slows down and under certain conditions its state of polarization changes due to light-matter interaction.  The speed of light through a medium (v) is specific to the density, refractive index and composition of the medium.  Several real life dynamic or static activities \cite{KTAGA, GAO, XIA1} modulate the refractive index (n) and density of the medium, which help us understand the physics and engineer various devices.   
   Light waves are transverse waves consisting of varying electric (E) and magnetic (B) fields that oscillate perpendicular to the direction of propagation and perpendicular to each other where the magnitude of electric field (E) is c (speed of light) times higher than that of the magnetic field (B).  The oscillating electric field vector on the transvers plane can be filtered along a particular direction by a polarizer. Time dependent rotation of the linearly polarized electric field vector of light due to dynamic modulation of refractive index and density of the medium as a result of the burning of fuel-air mixtures has been investigated here. We demonstrated that the acoustic waves and the moving flame front can be detected with an open single optical beam based polarimetry technique. In the past, optical science communities used the light-matter interaction with linearly polarized electric field vector to study various physical phenomena  \cite{KTAGA, GAO, XIA1, JIN} occurred in medium.  Here, we have used the principle of cross polarized field vector along Z-scan and leakage of depolarized light field vector through the analyser due to polarization vector rotation under the influence of dynamic modulation of density and refractive index during the passage of the flame front and the acoustic waves, as shown in Fig. 1.
\begin{figure}
\centering\includegraphics[width=12.98cm]{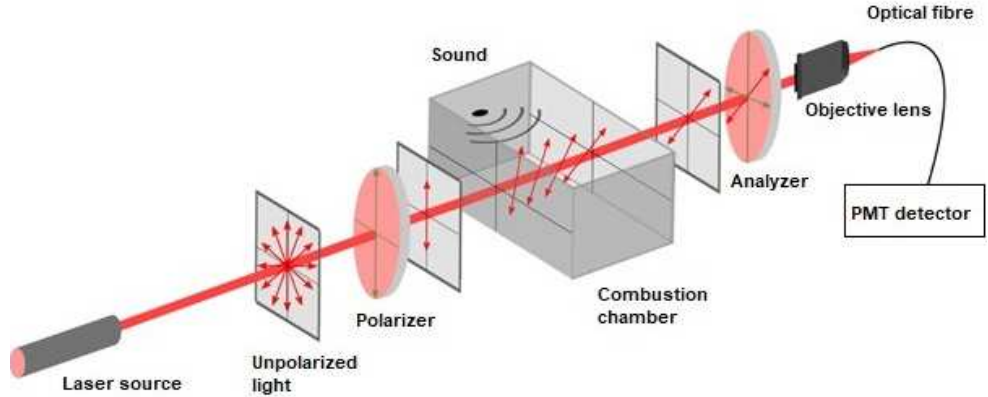}
\caption{Optical set up with the laser source, polarizer, analyser, objective 
lens, PMT and the combustion chamber where the acoustic and flame front are 
interacting with the electric field vector of the polarized light.  Leaked 
depolarized light is seen through the analyser due to the polarization vector 
rotation under the influence of acoustic waves and the moving flame front.}
\label{fig:pol}
\end{figure}
\section{Experimental facility}
The experimental facility \cite{SIBA} essentially consist of a channel of 400 mm length and 40×40 mm2 cross section with transparent walls (made of 8-15 mm thick acrylic plates), mass flow meters, storage tank for gaseous propane, air compressor, high speed camera (Phantom 7.3), ignition system, and a desktop PC for controlling the equipment. The end wall opposite to the ignition side is either closed by same acrylic plate (closed case) or by cellophane papers (in half open case) which ruptures when the pressure inside the channel increases more than a few atmosphere during combustion and flame propagation and saves unintended damage to the channel. A schematic of the facility is shown in Fig. 2 below. Here DDG stands for Digital Delay Generator, LPG for Liquid Petroleum Gas, and PMT for Photo Multiplier Tube. A photograph of the experimental set up with optical detection system in place is shown in Fig. 3. The various components of the optical system is also marked in the photo. The laser beam crosses the chamber span-wise. The circular reddish spots in Fig. 3 are the scattered laser light from the polariser, from the channel side walls, and from the head of the optical fibre mounted on the other side of the chamber to collect the refracted laser light. 

A collimated continuous wave laser beam (660 nm wavelength, 30 mW power) falls on the linear polarizer (working wavelength range 450-750 nm, surface flatness $\leq$ 4$\lambda$ at 633 nm, extinction ratio 4000:1, Newport Corporation) first. The circular red spots seen in Fig. 3 are the laser beam falling onto the polariser, and then onto the channel side walls. The polarized laser beam then passes through the combustion chamber (Fig. 2) span wise. On the opposite side of the chamber, an analyser is placed to block the unaffected polarized light to the sensing system. A 10X objective lens of numerical aperture 0.25NA and a 200 micron multimode bare optical fiber from Thorlabs (FT200UMT - 0.39 NA, $\phi$ 200 $\mu$m Core Multimode Optical Fiber, High OH for 300 - 1200 nm, TECS Clad) is used to collect the leaked depolarized laser light from the analyser. The other end of the optical fiber is connected to the input of a photomultiplier tube (assembled in-house using Hamamatsu R928 photomultiplier tube with a housing mount for PMT- K218). Finally the output signal from PMT is fed to an externally triggered oscilloscope to capture the signal as the acoustic disturbances and/or the flame front crosses the laser beam.

A schematic representation of the possible position of the pressure/acoustic waves and the flame front at three different instants of time is shown in Fig. 4. This is obtained from numerical simulation with hydrogen-air as the fuel air mixture, the details of which is given elsewhere and beyond the scope of this manuscript. The present experiments use propane-air as the reacting mixture at an equivalence ratio of 1.05. The position of the flame front and acoustic waves are assumed to be similar for the two fuel-air combinations. Deviations in the flow-field structure can happen as the flame velocity in hydrogen air mixture is higher than that of propane-air. The laser beam is placed perpendicular the plane of paper/side walls (Fig. 2) at various axial locations in various experimental realisations. The same is also represented schematically in Fig. 4. The laser beam was placed at five different axial positions from the ignition source namely, 3.75 cm, 11. 5 cm, 15 cm, 16.5 cm, and 18.7 cm while the ignition source was placed at a distance of 5 cm from the left end wall (please refer to Fig. 2). The vertical position of the beam is not important here as the objective of the present study is to detect the transversely oscillating acoustic waves and axially travelling flame front/acoustic wave (please refer to Fig. 4).
\begin{figure}
\centering\includegraphics[width=12.98cm]{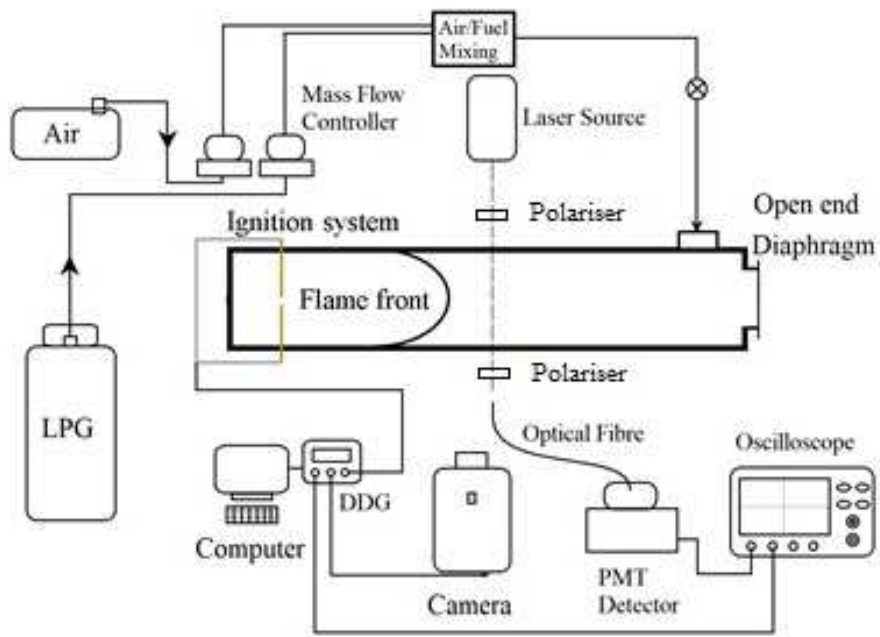}
\caption{A schematic of the experimental setup} 
\label{fig:exf1}
\end{figure}
\begin{figure}
\centering\includegraphics[width=12.98cm]{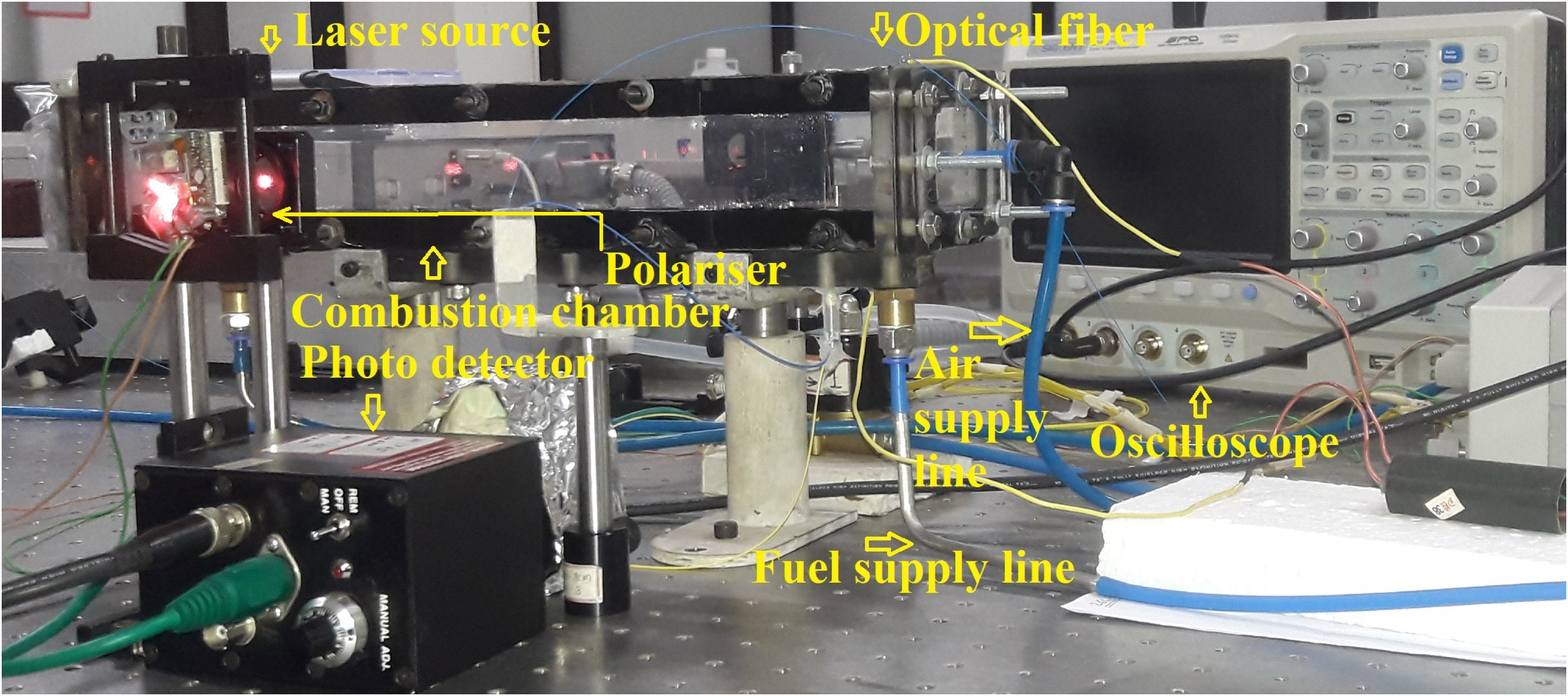}
\caption{Photograph of the experimental facility} 
\label{fig:exf2}
\end{figure}
\begin{figure}
\centering\includegraphics[width=12.98cm]{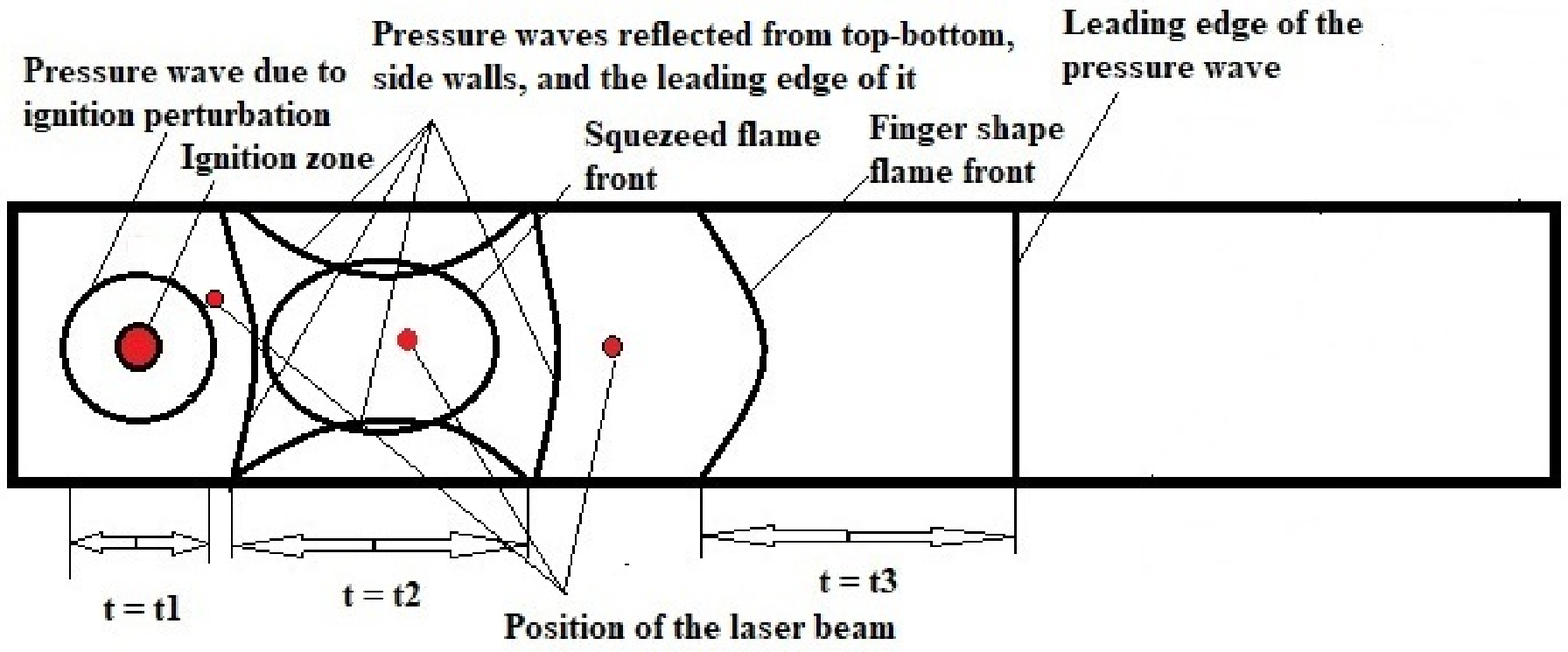}
\caption{Photograph of the experimental facility} 
\label{fig:exf3}
\end{figure}
\section{Results and discussions}
A set of photographs taken by a high speed camera (Phantom 7.3, Vison research) during the propagation of the flame front from the early stage immediately after the ignition till the inversion of the flame is shown in Fig. 5. The electrodes are also visible (a dark vertical line) near the right end of the channel. The signal obtained from the oscilloscope for the axial positions of the laser beam mentioned above is shown in the Fig. 6-9 below.  The reason for changing the axial position of the laser beam is to compare the oscilloscope signals, as the acoustics behaves and interacts with the flame front differently at different axial locations. There is little uncertainty over the timing of the spark. However, when the signal from the oscilloscope and photographs captured by camera are compared in various realizations, the acoustics and flame front are identified. When the light source is placed very close to the electrodes (3.75 cm), the signal captured by the oscilloscope is mainly because of the acoustic waves (Fig. 6). A weak and developing flame front is visible in the oscilloscope signal. At a distance of 11.5 cm (Fig. 7), the oscilloscope signal shows the presence of strong acoustic waves. The strength of the acoustic waves gradually decreases over time as flame front has already crossed it and there is no supply of heat energy to it. Clear cut crossing of the flame front over the light source, flattening of the flame front and tulip inversion is clearly discernible when the light source is placed at 15 cm downstream of the ignition source (Fig. 8(top)). At further downstream location of the light source (16.5 cm, Fig. 8(bottom)), the effect of the flame on the light source becomes less than the previous position because the acoustic wave and flame front reach the light source almost simultaneously. At a distance of 18.7 cm of the light source, the effect of both flame front and acoustics on it are visible (Fig. 9). The flame flattening and inversion happen behind this position. When the light source is obstructed, only the triggering signal is visible (Fig. 10) in channel 1 of the oscilloscope and no signal is obtained in channel 2, which is fed by signal from the photodetector. This confirms that the oscilloscope signals shown in Fig. 6-9 are real and not due to electrical noise.

\begin{figure}[htb]
\centering\includegraphics[width=12.98cm]{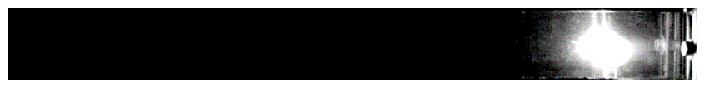}
\centering\includegraphics[width=12.98cm]{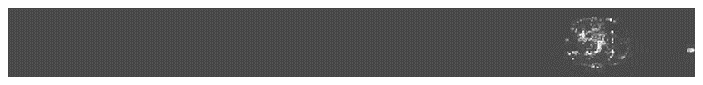}
\centering\includegraphics[width=12.98cm]{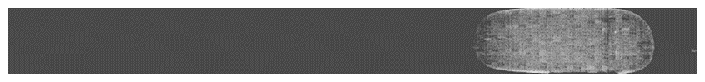}
\centering\includegraphics[width=12.98cm]{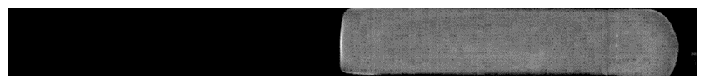}
\centering\includegraphics[width=12.98cm]{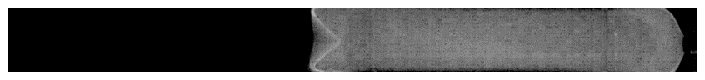}
\caption{Partial snapshots of the flow-field during the propagation of the flame front from the ignition source till the inversion of the flame, (from top to bottom),  Sparking, t = 0 ms, elliptical flame front at t = 8.5 ms, finger shape flame at t = 16.75 ms, flame flattening at t = 26.25 ms, Tulip inversion at 31.75 ms} 
\label{fig:snp}
\end{figure}
\begin{figure}[htb]
\centering\includegraphics[width=10.0cm]{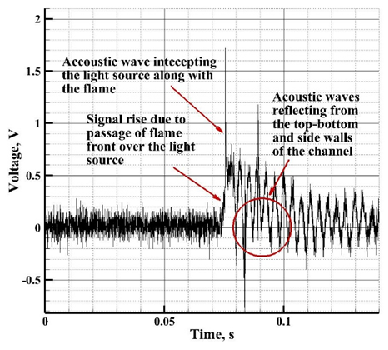}
\centering\includegraphics[width=10.0cm]{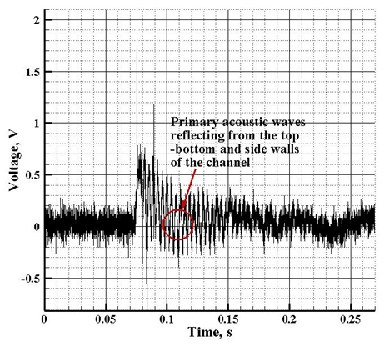}
\caption{Oscilloscope signal when the light source was placed at a distance of 3.75 cm downstream of the spark location (top) zoomed in (bottom) full signal history.} 
\label{fig:6}
\end{figure}
\begin{figure}[htb]
\centering\includegraphics[width=10.0cm]{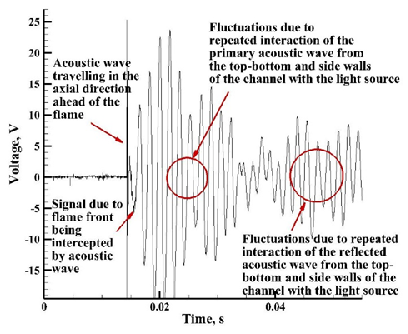}
\centering\includegraphics[width=10.0cm]{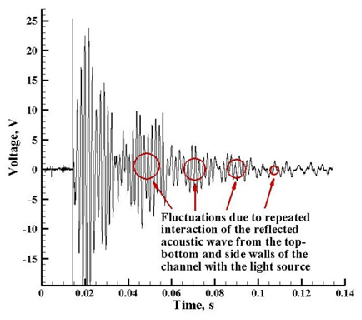}
\caption{Oscilloscope signal when the light source was placed at a distance of 11.5 cm downstream of the spark location (top) zoomed in (bottom) full signal history.} 
\label{fig:7}
\end{figure}
\begin{figure}[htb]
\centering\includegraphics[width=10.0cm]{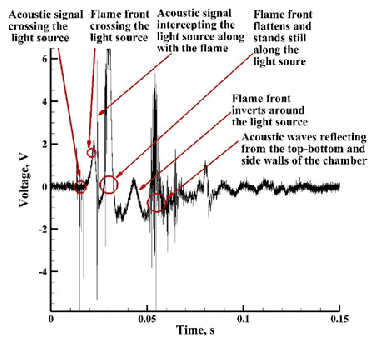}
\centering\includegraphics[width=10.0cm]{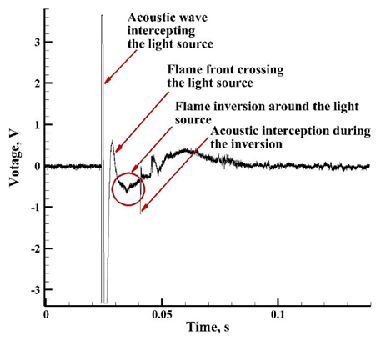}
\caption{Oscilloscope signal when the light source was placed at a distance of (top) 15 cm (bottom) 16.5 cm downstream of the spark location} 
\label{fig:8}
\end{figure}
\begin{figure}[htb]
\centering\includegraphics[width=10.0cm]{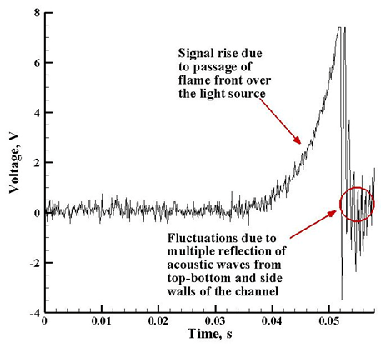}
\centering\includegraphics[width=10.0cm]{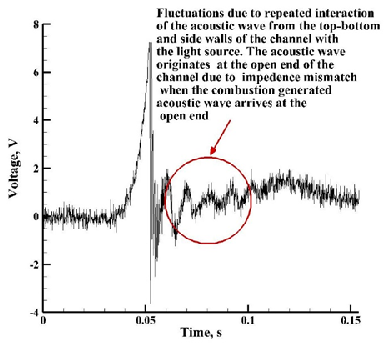}
\caption{Oscilloscope signal when the light source was placed at a distance of 18.7 cm downstream of the spark location (a) zoomed in (b) full signal history.} 
\label{fig:9}
\end{figure}
\begin{figure}[htb]
\centering\includegraphics[width=10.0cm]{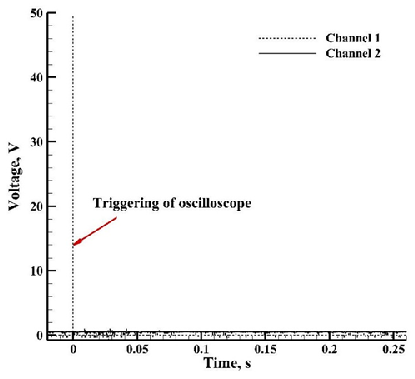}
\caption{Oscilloscope signals when the light source was obstructed and placed at a distance of 18.7 cm downstream of the ignition source. Channel 1 represents the triggering signal of the oscilloscope, and channel 2 represents the signal obtained from the photodetector.} 
\label{fig:9}
\end{figure}
\section{Conclusions}
An optical technique based on the principles of polarization and depolarization effect under lighter-matter interaction phenomenon is used to capture thermo-acoustics disturbances, and a moving premixed flame front simultaneously. Photographs taken by a high speed camera and signals obtained from oscilloscope are compared to identify the flame front position and its interception by the acoustic waves at various axial positions and instants of time. The results are consistent with the numerical predictions, and can also be justified physically. Future study would focus on calibration of the acoustic signal by a known impulsive pressure wave. The technique could be applied in locating and measuring (after calibration) the complex thermo-acoustic disturbances arising in modern gas turbine combustors.  


\begin{thebibliography}{9}
\bibitem{ELL}            
O.C. Ellis, R.V. Wheeler,   
CCCCXXVI, Explosions in closed cylinders. Part III. The manner of movement of flame.  
\textit{J. Chem. Soc.}1928; 
\textbf{0}:3215--3218.

\bibitem{SAL}    
G.D. Salamandra, T.V. Bazhenova, I.M. Naboko,             
Formation of detonation wave during combustion of gas in combustion tube.    
\textit{Symp. (Int.) Combust. }1958; 
\textbf{7}(1):851--858.

\bibitem{DUN}    
D. Dunn-Rankin, R.F. Sawyer,
Tulip flames: changes in shape of premixed flames propagating in closed tubes.
\textit{Expt. Fld.}1998; 
\textbf{24}:130--140.

\bibitem{HUA}    
Huahua Xiao, Ryan W. Houim, Elaine S. Oran,
Effects of pressure waves on the stability of flames propagating in tubes.
\textit{Proc. Combust. Inst.}2017; 
\textbf{36}(1):1577--1583.

\bibitem{XIA}   
Xiaoxi Li, Huahua Xiao, Qiangling Duan, Jinhua Sun,
Numerical study of premixed flame dynamics in a closed tube: Effect of wall boundary condition.  
\textit{Proc. Combust. Inst.}2021; 
\textbf{38}(2):2075--2082.

\bibitem{SIBA}   
Siba Prasad Choudhury, Ratan Joarder, 
High-speed photography and background oriented schlieren techniques for characterizing tulip flame.  
\textit{Combust. Flame }2022; 
\textbf{245}.

\bibitem{KTAGA}  
K. Taga, R. Hisatomi, Y. Ohnuma, 
Optical polarimetric measurement of surface acoustic waves. 
\textit{Appl. Phys. Lett. }2021; 
\textbf{119}(181106).

\bibitem{GAO}  
Gaoshang Liu, Jichuan Xiong, Yun Cao, Ruijie Hou, Lishan Zhi, Zhiying Xia, Weiping Liu, Xuefeng Liu, Christ Glorieux, John H. Marsh, and Lianping Hou,  
Visualization of ultrasonic wave field by stroboscopic polarization selective imaging. 
\textit{Opt. Exp.}2020; 
\textbf{28}(18):27096--27106.

\bibitem{XIA1} 
Xiaoyi Zhu, Zhiyu Huang, Guohe Wang, Wenzhao Li, Da Zou, and Changhui Li,
Ultrasonic detection based on polarization-dependent optical reflection.  
\textit{Opt. Lett.}2017; 
\textbf{42}(3):439--441.

\bibitem{JIN}
Jingfei Yin, Qian Bai, Bi Zhang, 
Sensitivity of polarized laser scattering detection to subsurface damage in ground silicon wafers. 
\textit{Mat. Sc. Semicon. Proc.}2022; 
\textbf{144}, 106570.

\end{thebibliography}
\end{document}